\documentclass[aps, prl, reprint]{revtex4-2}  

\usepackage[english]{babel}
\usepackage{bm}
\usepackage{graphicx}
\usepackage{float}
\usepackage{amsmath,amssymb}
\usepackage{physics}
\usepackage{hyperref}
\usepackage{xcolor}

\newif\ifhighlight

\newcommand{\highlight}{\highlighttrue}
\highlight

\newcommand{\editor}[2]{%
  \expandafter\newcommand\csname #1note\endcsname[1]{%
    \textcolor{#2}{(\textbf{#1:} ##1)}}%
  \expandafter\newcommand\csname #1\endcsname[1]{%
    \ifhighlight\textcolor{#2}{##1}\else##1\fi}%
  \expandafter\newcommand\csname #1cancel\endcsname[1]{%
    \ifhighlight\textcolor{#2}{\sout{##1}}\fi}%
  \expandafter\newcommand\csname #1change\endcsname[2]{%
    \ifhighlight\textcolor{#2}{\sout{##1} ##2}\else##2\fi}%
  \newenvironment{#1text}{\ifhighlight\color{#2}\fi}{\color{black}}%
}

\editor{resub}{blue}

\begin{document}

\title{Exact {Electrostatic} Theory of Fermi-Energy Response at Metallic Interfaces}

\author{Th\'eophane Bernhard}
\affiliation{Physicochimie des \'Electrolytes et Nanosyst\`emes Interfaciaux, Sorbonne Universit\'e, CNRS, F-75005 Paris, France}

\author{Andrea Grisafi}
\email{andrea.grisafi@sorbonne-universite.fr}
\affiliation{Physicochimie des \'Electrolytes et Nanosyst\`emes Interfaciaux, Sorbonne Universit\'e, CNRS, F-75005 Paris, France}

\begin{abstract}
    The response of the Fermi energy to external perturbations governs key physical observables at metallic interfaces. Although this response admits a local formulation in terms of the Fukui function, its evaluation has traditionally been limited by inherent approximations, fundamentally rooted in the difficulty of adding a finite charge in a periodic system. We present an exact  resolution to this problem that{, within an electronic-structure framework capturing metallic behavior,} exploits the screening properties of electronic conductors to compute Fukui functions via a finite electric~field. The resulting linear-response theory yields strictly quadratic error scaling of Fermi-level shifts across representative platinum surfaces, achieving sub-meV accuracy up to fields of 0.1~V/\AA. The approach is further validated by reproducing work-function changes under molecular perturbations, and by providing mean-field estimates of electrode potentials that yield capacitance--voltage curves consistent with experiment. Our findings establish a rigorous foundation for a local theory relating electrostatic screening and Fermi-energy variations at metallic interfaces. 
\end{abstract}

\maketitle 

Positioning the Fermi level at metallic interfaces plays a central role in the calculation of numerous physical properties. Notable examples include 
work functions in functional materials~\cite{DeBoer2005,Lin2023,Schultz2023,Niederreiter2023}, electrode potentials in electrochemical devices~\cite{Haruyama2018,Le2021,Melander2024}, gate voltage effects in metal/semiconductor heterojunctions~\cite{Majumdar1995,Anzi2022,Chiu2024}, and onset potentials in electrocatalytic reactions~\cite{Xiao2016,Che2018,Batchelor-McAuley2023}. 
In all these cases, a key challenge is to determine how a variation of external potential shifts the Fermi level of the metal with respect to a well-defined energy reference. This response fundamentally arises as a consequence of interfacial screening, producing a uniform electrostatic potential shift in the metallic bulk that rigidly offsets the electronic chemical potential.  

In principle, an explicit formulation of the  Fermi-level response is provided through an extension of density functional theory (DFT), commonly referred to as ``conceptual'' DFT~\cite{Li1995,Geerlings2003}. 
In the present case, conceptual-DFT allows us to establish a local linear-response theory for the variation of the electronic chemical potential, namely the metal Fermi level $E_\text{F}$, induced by a relatively small variation of external potential $v_\text{ext}(\mathbf{r})$:
\begin{equation}\label{eq:fermi-response}
    \Delta E_\text{F} = \int \mathrm{d}\mathbf{r}\, f(\mathbf{r})\, \Delta v_\text{ext}(\mathbf{r}) + \mathcal{O}(\Delta v_\text{ext}^2)\, .
\end{equation}
The linear-response functions $f(\mathbf{r})$ are known as Fukui functions~\cite{Parr1984,Fukui1975}. Although their definition as the first functional derivative of $E_\text{F}$ with respect to $v_\text{ext}(\mathbf{r})$ makes the evaluation cumbersome, a Maxwell relation is commonly adopted to enable a more convenient rewriting in terms of derivative of the electron density $\rho(\mathbf{r})$ with respect to the number of electrons $N$~\cite{Parr1984}: 
\begin{equation}\label{eq:fukui-def}
    f(\mathbf{r})=  \left(\frac{\delta E_\text{F}}{\delta v_\text{ext}(\mathbf{r})}\right)_{N} = \left(\frac{\partial \rho(\mathbf{r})}{\partial N}\right)_{v_\text{ext}(\mathbf{r})}\, .
\end{equation}
Importantly, this definition becomes unambiguous for the exclusive case of electronic conductors, where $N$ can be continuously varied across the Fermi level without abrupt changes of the electronic structure~\cite{Perdew1982,Yang1985,Yang2012,Morgante2023}. 

When dealing with periodic condensed-phase systems, performing the calculation of Fukui functions presents a particular challenge~\cite{Sablon2007,Allison2013,Barrera2025}. The key difficulty lies in estimating the derivative of the electron density by finite differences, as any variation of the number of electrons necessarily introduce a net charge in the system. This poses a well-known hurdle to periodic DFT calculations, where the excess/defect of electronic charge in the simulation cell must be somehow perfectly compensated to avoid a divergence of the electrostatic energy. The most common strategy to enforce global electroneutrality is a homogeneous background of charge (HBC) that is implicitly added in the calculation whenever adopting a plane-waves representation of the charge density. However, the HBC introduces an unphysical polarization of the cell that dramatically affects the Fukui function calculation~\cite{Barrera2025}. Notable attempts to mitigate this problem have been developed over the years, ranging from general correction schemes of periodic DFT calculations~\cite{Jarvis1997,Krishnaswamy2015,Silva2021}, to methods that are specifically tailored to treat the charging of a metal surface~\cite{Filhol2006,Taylor2006,Mamatkulov2011,Filhol2014}.

A different possibility is that of relying on the relation between Fukui functions and the local density of states (LDOS) at $E_\text{F}$, as derived from Kohn-Sham and other single-particle representations~\cite{Yang1985}. Although this strategy  bypasses the problem of adding a finite charge in the system, major difficulties arise in properly accounting for the relaxation of the electronic states around the Fermi level~\cite{Cohen1994,Cardenas2008,Ceron2020}. A recent study performed an extensive comparison between LDOS and finite-charge methods for computing Fukui functions in various metals and semiconductors, highlighting advantages and disadvantages of the different approaches depending on the system at hand~\cite{Barrera2025}. By and large, no existing method can provide an exact solution to this problem. In what follows, we show that perfect electrostatic screening in electronic conductors enables an alternative calculation of Fukui functions at metallic interfaces that is free of the aforementioned approximations.

We start by considering an electronically conductive material that is periodically repeated along the $xy$ plane, thus exposing two surfaces along the nonconductive direction $z$. No specific assumption is made about the symmetry of the two surfaces, which can be clean, defective, or present an interface with an electronic insulator. From Eq.~\eqref{eq:fukui-def}, Fukui functions can in principle be computed by finite differences under a (possibly fractional) electron addition and removal, $\pm\Delta N$.
From Gauss theorem, the charge-density variations derived from this process must be localized at the surface of the electronic conductor, i.e., $\Delta \rho(\mathbf{r})$ must vanish in the metallic bulk. Then, it makes sense to partition the system in a left ($z<0$) and right ($z>0$) contribution, by setting $z=0$ in the bulk. Specifically, we can split the variation of electronic charge as $\Delta N = Q_\text{L} + Q_\text{R}$, where $Q_\text{L/R}=\int_{z \lessgtr 0}\mathrm{d}\mathbf{r}\, \Delta \rho(\mathbf{r})$.

We now make the key observation that, from the perspective of the metallic bulk, the distribution of electronic charge accumulated at the two surfaces is viewed as a homogeneous surface charge density, namely $\sigma_\text{L} = Q_\text{L}/A$ and $\sigma_\text{R}= Q_\text{R}/A$, respectively, with $A$ the $xy$ surface area of the periodic cell. This will always be true for sufficiently thick metal slabs, where the inhomogeneities of the charge distributions are smeared out from a sufficiently large distance. In this limit, the electric field at $z=0$ can be written without loss of generality as $E(0) = 2\pi \sigma_\text{L} -2\pi \sigma_\text{R} = 0$. Regardless from the asymmetry of the two surfaces, the charge injected in the system must then be equally partitioned between the left and right hand side, i.e., $Q_\text{L} = Q_\text{R} = \Delta N/2 \equiv Q$. %
Therefore, we obtain a new definition of Fukui functions, where the variations of the electron density are measured with respect to the corresponding amount of surface charge:
\begin{equation}\label{eq:fukui-surface}
    f(\mathbf{r}) = \lim_{\Delta N\to 0} \frac{\Delta \rho_{+\Delta N}(\mathbf{r})-\Delta \rho_{-\Delta N}(\mathbf{r})}{4\,  Q_{\Delta N}} = \frac{1}{2}\frac{\partial \rho(\mathbf{r})}{\partial Q}\, .
\end{equation}

The result just derived suggests that Fukui functions are fundamentally local surface properties of electronic conductors, so that the  interfacial response is insensitive to the global mechanism by which the surface charge $Q$ is generated.  Based on this insight, we make the hypothesis that the charge accumulation at any given surface could be independently induced by an applied electric field $\varepsilon_z$ along the $z$ direction, thus considering the complementary side of the slab as a virtually infinite reservoir of electrons. Namely, we assume that we can rewrite Eq.~\eqref{eq:fukui-surface} using the following chain rule:
\begin{equation}\label{eq:fukui-field}
      f(\mathbf{r}) = \frac{1}{2}\frac{\partial \rho(\mathbf{r})}{\partial \varepsilon_z}\frac{\partial \varepsilon_z}{\partial Q} = 
      \lim_{\varepsilon_z \to 0} \frac{\Delta \rho_{+\varepsilon_z}(\mathbf{r})-\Delta \rho_{-\varepsilon_z}(\mathbf{r})}{4\, Q_{\varepsilon_z}}\, . 
\end{equation}
This time, the applied field has the effect of polarizing the metal, yielding an antisymmetric charge distribution that perfectly screens $\varepsilon_z$ in the metallic bulk.
Since left and right surfaces have opposite charges, the two interfaces must now be separately treated when applying Eq.~$\eqref{eq:fukui-field}$ to compute Fukui functions. 

\begin{figure}[t]
    \centering
    \includegraphics[width=0.9\linewidth]{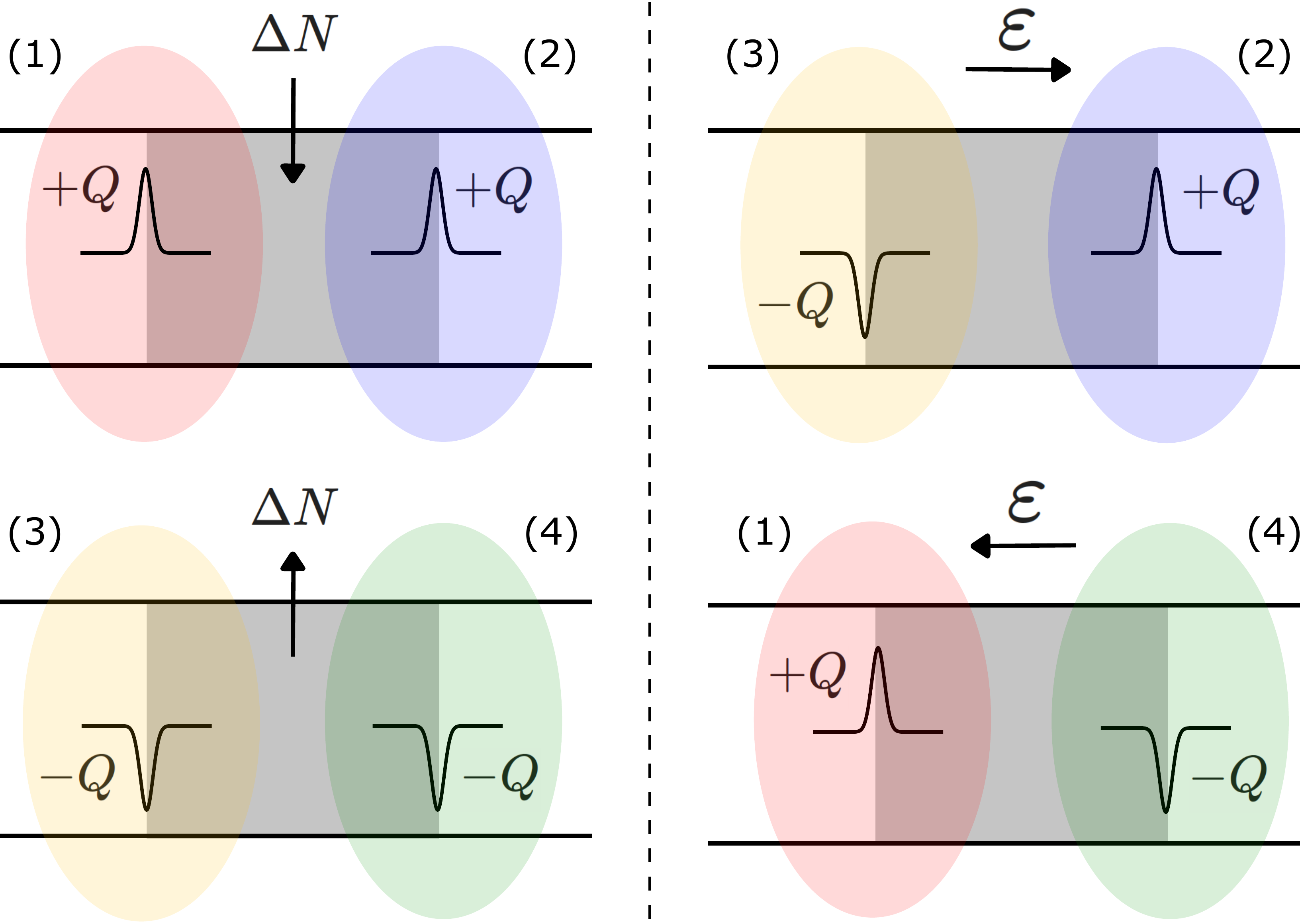}
    \caption{Comparison between finite-charge ($\pm \Delta N$) and finite-field ($\pm \varepsilon_z$) methods for computing Fukui functions at the interface between an electronic conductor (\textit{gray area}) and an electronic insulator (\textit{white area}). Highlighted regions of equivalent color and number label indicate physically analogous interfaces. No spatial symmetry of the electronic charge distributions is assumed between left and right surfaces.}
    \label{fig:method}
\end{figure}

To show that Eqs.~\eqref{eq:fukui-surface} and~\eqref{eq:fukui-field} can yield equivalent results, let us compare the two physical scenarios where finite-charge  $(\pm \Delta N)$ and finite-field $(\pm \varepsilon_z)$ perturbations are applied to the system; these are illustrated in Fig.~\ref{fig:method}.  As a consequence of perfect screening, it is always possible to choose the value of the applied field such that the highlighted pairs of metallic interfaces appear as subject to the same external potential, and, as such, are physically indistinguishable from each other. This remains true even if the system was periodically repeated along the  $z$ direction, as any added linear polarization term 
can be formally absorbed into $\varepsilon_z$. 
As the field value enters only parametrically through the chain rule, it follows that each pair of physically analogous interfaces depicted in Fig.~\ref{fig:method} provides a formally equivalent contribution to the finite-charge and finite-field formulations of Fukui functions. We refer to the Supplemental Material (SM), Ref.~\cite{suppmat}, for a thorough demonstration. 
Despite the formal correspondence, the finite-field pathway carries the great practical advantage that the system remains electroneutral, thereby sidestepping any HBC artifact associated with conventional finite-charge approaches. 

\begin{figure}[t]
    \centering
    \includegraphics[width=1.0\linewidth]{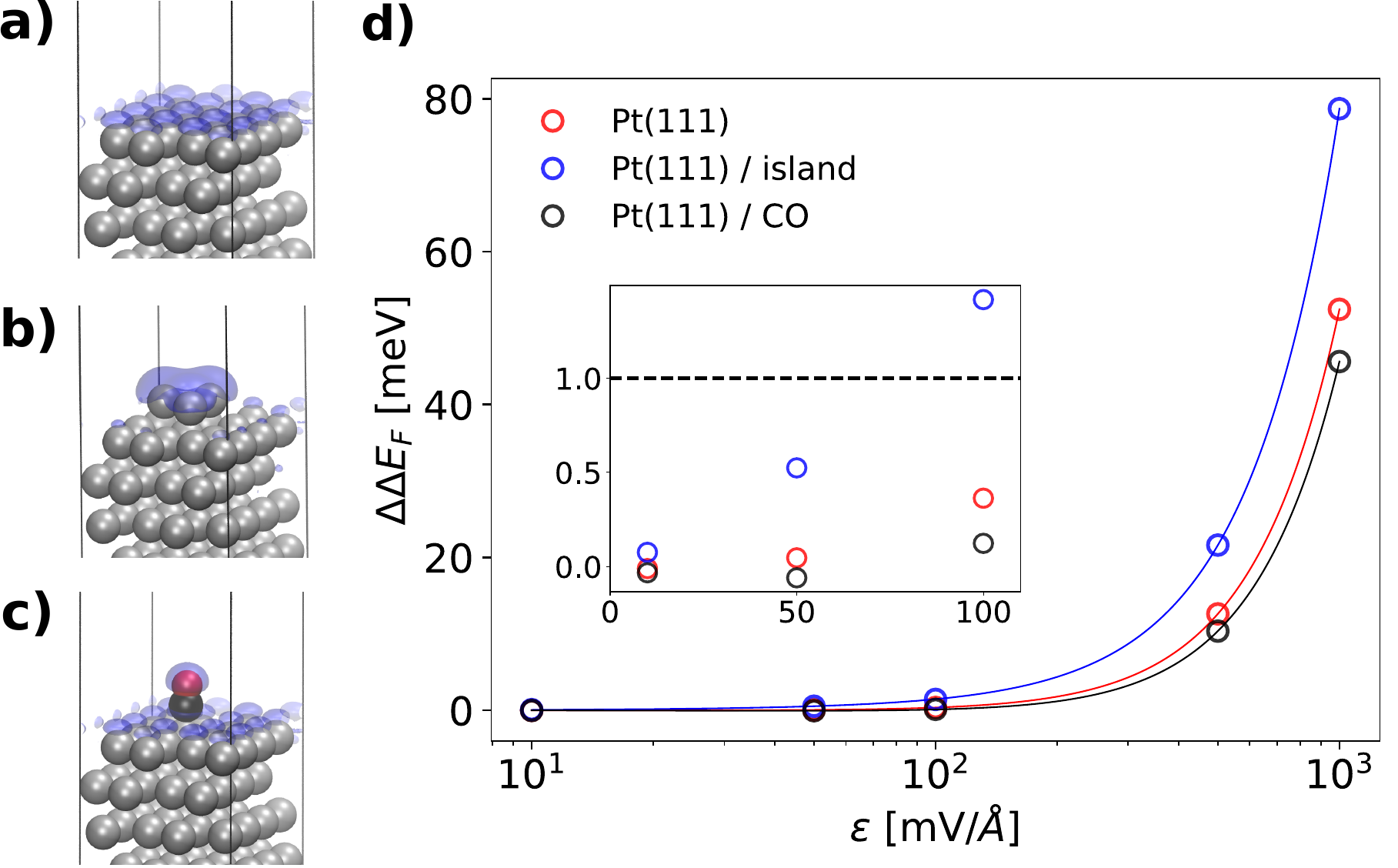}
    \caption{Computed Fukui functions for three relaxed Pt(111) surfaces, reported as 3D isovalues: a) bare metal slab, b) metal slab with a 3 Pt-atoms island, c) metal slab with an adsorbed CO molecule on a top Pt-site. 
    d) Signed error of Fermi energy variations $\Delta E_\text{F}$ computed through our linear-response theory, for systems subject to linear potentials $v(z) = -\varepsilon_z z$ of increasing electric field $\varepsilon_z$.  Full lines are obtained from a monomial quadratic fit of the data. Inset: magnified view for $\varepsilon_z\le0.1$ V/{\AA}, showing an overall sub-meV accuracy.}
    \label{fig:fukui_pt}
\end{figure}

We begin by testing our theory on three representative Pt(111) surfaces, namely the bare slab (Fig.~\ref{fig:fukui_pt}-a), a defective slab with a three Pt-atoms island (Fig.~\ref{fig:fukui_pt}-b), and a slab with a CO molecule adsorbed on a top Pt-site (Fig.~\ref{fig:fukui_pt}-c). Finite-field calculations of Fukui functions on fully relaxed geometries are performed as in Eq.~\eqref{eq:fukui-field}. {In particular, we observe that as few as seven platinum layers are enough to achieve perfect electrostatic screening, as  required from the assumed localized response of the electron density at the metal surface.} Additional computational details 
are reported in the SM~\cite{suppmat}. 

From Eq.~\eqref{eq:fermi-response}, the Fukui functions so computed can be used to determine the Fermi-energy variation, $\Delta E_\text{F}$,   under external potentials.  For that, we expose the three platinum surfaces to a linear potential profile along $z$, i.e., $v_\text{ext}(z) = -\varepsilon_z z$, thus using $\varepsilon_z$ as a parameter to tune the magnitude of the applied perturbation. Results are compared against reference DFT values of Fermi-level shifts computed at the same level of theory adopted to perform the Fukui functions calculations. The signed error, $\Delta \Delta E_\text{F}$, of our linear-response predictions is reported in Fig.~\ref{fig:fukui_pt}-d for increasing values of~$\varepsilon_z$. Notably, we observe an error that is generally lower than 1~meV for $\varepsilon_z=0.1$~V/{\AA}, and that remains below 0.1~eV for large fields of $\varepsilon_z=1.0$~V/{\AA}. From a monomial quadratic fit of the data, depicted by the full lines in Fig.~\ref{fig:fukui_pt}-d, we find that the leading contribution to the error exhibits a strictly quadratic growth with the magnitude of the applied field, i.e., $\Delta \Delta E_\text{F} \sim\mathcal{O}(\varepsilon_z^2)$. As no residual linear source of error appears in our predictions, this result provides rigorous numerical evidence about the exactness of a linear response theory based on the Fukui functions computed via our approach. {We note that the term exact is not meant in a universal sense, but rather refers to the absence of additional approximations beyond the chosen level of electronic-structure theory that correctly captures metallic screening. In practice, semilocal density-functionals like PBE~\cite{perd+96prl} are shown to serve this scope well, while hybrid density-functional approximations can lead to small spurious bulk responses due to localization of the electronic states~\cite{suppmat}.}

\begin{figure}[b]
    \centering
    \includegraphics[width=0.9\linewidth]{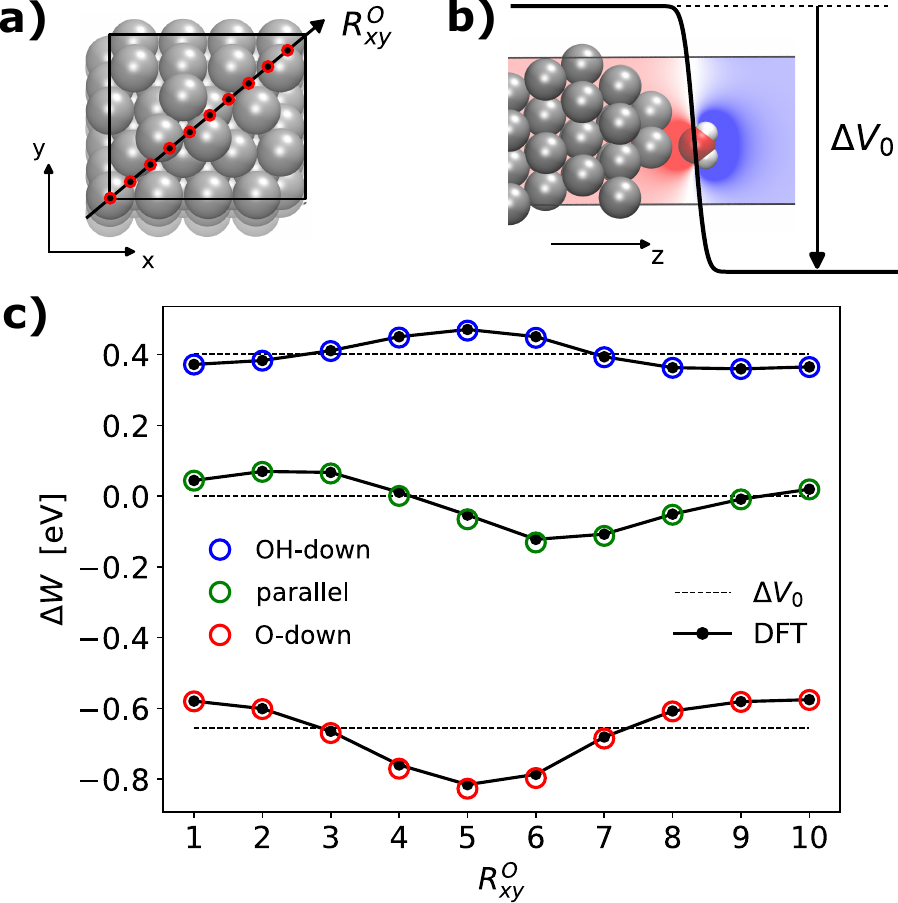}
    \caption{a) Representation of ten lateral displacements $R^\text{O}_{xy}$ over the $xy$-plane of a water molecule at 3~{\AA} distance from a defective Pt(111) surface. b) Dipolar potential drop $\Delta V_0$ associated with the planar $xy$-average of the molecular external potential, $\Delta \bar{v}_\text{ext}(z)$, for an example upright orientation. 
    c) Empty circles: $\Delta W$ response induced by the field of three molecular orientations. Black dots and lines: DFT reference. Black dashed lines: external potential drops $\Delta V_0$.  }
    \label{fig:lateral-fermi}
\end{figure}

An important application of the theory, particularly relevant in surface measurements, involves the determination of work-function changes~\cite{Lin2023}. These are defined as $\Delta W= \Delta V_\text{vac}-\Delta E_\text{F}$, where $\Delta V_\text{vac}$ denotes the shift of electronic vacuum level. For a charge-neutral perturbation localized near the surface, the induced electronic response produces, to leading order, a dipolar charge distribution. Upon adopting a 2D-periodic electrostatic solver~\cite{Minary2002}, a flat vacuum level can be defined, giving rise to a step in the planar-averaged electrostatic potential across the interface equal to $\Delta W$. The height of this step is often interpreted via the Helmholtz relation~\cite{Lang1971}, which connects the difference between $\Delta V_\text{vac}$ and the potential variation in the metallic bulk ($\Delta V_\text{bulk}\simeq \Delta E_\text{F}$) to the interfacial polarization per unit area, i.e., $\Delta W = - 4\pi p_z/A$. Because only potential differences are physically meaningful, the external potential can be referenced about its spatial average  without affecting the work-function calculation---a feature that is in fact automatically enforced from the vanishing $\textbf{G}=0$ component of $\Delta v_\text{ext}(\mathbf{r})$. This referencing here implies that the interfacial step variation of the electrostatic potential presents equal and opposite asymptotes, so that  $\Delta E_\text{F} \simeq \Delta V_\text{bulk}= - \Delta V_\text{vac}$. It follows that any gauge-invariant work-function change resulting from a charge-neutral perturbation can be directly computed from our linear-response theory as
\begin{equation}\label{eq:work}
    \Delta W = - 2 \int \mathrm{d} \mathbf{r}\, f(\mathbf{r}) \left(\Delta v_\text{ext}(\mathbf{r}) - \langle\Delta v_\text{ext}\rangle_\Omega \right)\, .
\end{equation}

As a prototypical example that lifts any symmetry constraint about both the surface and perturbation geometries, we consider the response of the defective Pt configuration depicted in Fig.~\ref{fig:method}-b under the external potential of different gas-phase water orientations (namely, upright O-down, OH-down and flat geometries) at a distance of 3~{\AA} from the nearest surface atoms. Eq.~\eqref{eq:work} is then used to compute $\Delta W$ under ten lateral displacements of the oxygen atom on the $xy$ plane (Fig.~\ref{fig:lateral-fermi}-a). From a direct comparison with reference DFT calculations, reported in Fig.~\ref{fig:lateral-fermi}-c, we observe how the derived work-function changes are predicted with extremely high accuracy in all three cases, including that associated with the flat-geometry quadrupolar perturbation. In particular, we note how the theory is capable of reproducing fluctuations of the order of 0.1~eV around the expected dipolar jump ($\Delta V_0$) that  trivially reflects the asymptotic variation of  the molecular external potential (Fig.~\ref{fig:lateral-fermi}-b). While the caveat of a fixed perturbation can be seen as a limiting factor when considering molecular adsorptions driven by chemical bonding, this result demonstrates the remarkable ability of the exact linear-response framework to quantitatively capture how the atomistic details of the surface influence the interfacial screening and derived work-function changes.

\begin{figure}[t]
    \centering
    \includegraphics[width=0.9\linewidth]{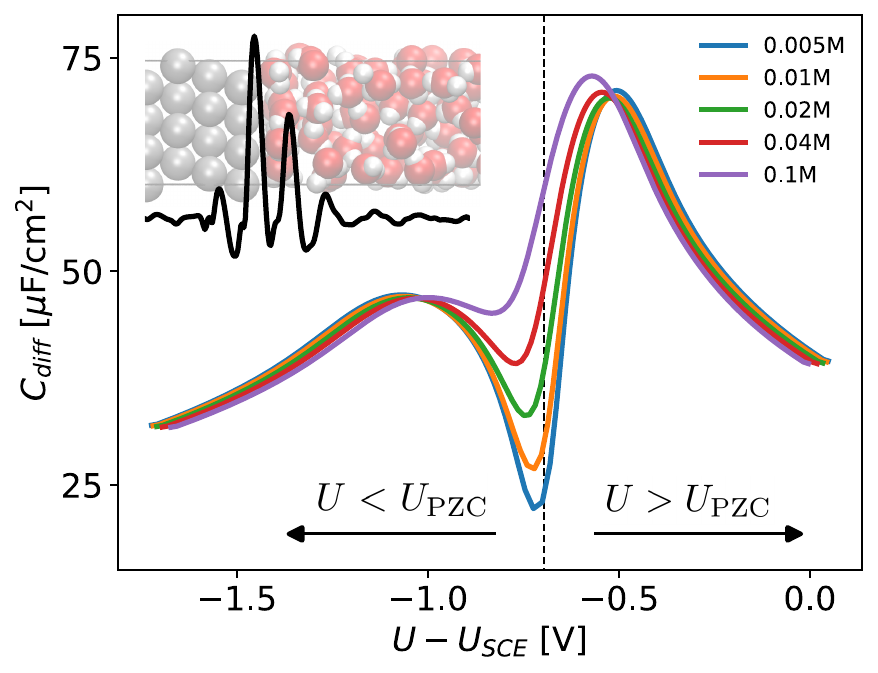}
    \caption{Computed capacitance ($C_\text{diff}$) versus potential ($U$) curves of the Ag(111)/NaF(aq) interface at increasing ionic concentrations. Predicted values of electrode potentials are shifted by the PZC value of $-0.695$ V (dashed line), used to compare directly with the experimental saturated calomel electrode (SCE) potential scale reported in Ref.~\cite{Valette1989}. Inset: planar integral of the thermally averaged Fukui function of the Ag(111)/water interface used as reference PZC state.
    }
    \label{fig:ag-water-fukui}
\end{figure}

Having validated the method under various surfaces and external perturbations, we proceed by considering the response of a full metal--insulator interface. As an example, we consider a Ag(111) slab in contact  with liquid water under 3D periodic boundary conditions. 
The finite-field calculation of Fukui functions  for the full metal/water interface is performed on a total of 400 independent frames selected every 0.5~ps from a classical simulation~\cite{suppmat}. The Fukui function planar integral is depicted in the inset of Fig.~\ref{fig:ag-water-fukui} as a statistical average over the simulated trajectory, $\langle \bar{f}(z)\rangle$.  We find an oscillatory profile that decays into the water bulk, exposing the interfacial nature of the response at thermal equilibrium via three main spillover peaks. This observation is consistent with recent \textit{ab initio} results about the charge-density response at electrified metal/electrolyte interfaces, revealing how a substantial portion of surface electronic charge leaks into the liquid phase~\cite{Andersson2025,Li2025}. 

We now aim to show that the computed response function can be used to estimate capacitance--voltage (C--V) curves at electrochemical interfaces---a paradigmatic example where the adoption of purely first-principle approaches is notoriously challenging~\cite{Jia-Bo2020,Li2025,Zhu2025}.  In this context, positioning the Fermi level plays a key role for the determination of the electrode potential, $U$. While the individual values of $U$ depend on the choice of reference electrode, the difference between~$U$ and the so-called potential of zero (free) charge,~$U_\text{PZC}$, represents an intrinsic property of the electrode/electrolyte interface~\cite{Li2022,Li2025}. In practice, $U-U_\text{PZC}$ can be computed from the thermal average of the Fermi-level shift, $\left<\Delta E_\text{F}\right>$, measured with respect to the equilibrium variation of the (physical) electrostatic potential $\phi$ in the liquid bulk~\cite{Le2017}:
\begin{equation}\label{eq:electrode-potential}
U-U_\text{PZC} = -\left<\Delta E_\text{F}\right>/e - \left<\Delta \phi_\text{L}\right>\, .
\end{equation}

We consider the case of the Ag(111)/NaF(aq) interface, whose C--V profiles can be directly compared with experimental measurements~\cite{Valette1989}.
At low concentrations of the NaF aqueous solution, the PZC state can be well approximated by the Ag(111)/water interface at zero surface charge~\cite{Le2017,Li2022}. This allows us to interpret the statistically averaged Fukui function, $\left<\bar{f}(z)\right>$, reported in Fig.~\ref{fig:ag-water-fukui} as the equilibrium response function of the system at the PZC, from which $\left<\Delta E_\text{F}\right>$ can be directly obtained as in Eq.~\eqref{eq:fermi-response} under a mean-field approximation. In this context, the perturbing potential can be defined from the equilibrium variation of the electrostatic potential profile $\left<\Delta \phi(z)\right>$ that is formed at the Ag(111)/NaF(aq) interface as a consequence of the electrode charging. Since $\left<\Delta \phi(z\to\infty)\right>\equiv\left<\Delta \phi_\text{L}\right>$ asymptotically far from the surface, this definition is internally consistent with the calculation of Eq.~\eqref{eq:electrode-potential}, making $U-U_\text{PZC}$ insensitive to the choice of potential reference.  
Here, we represent $\left<\Delta \phi(z)\right>$ through a modified Poisson-Boltzmann model~\cite{Kornyshev2007}, which is solved under fixed electrode surface charges, from $\sigma_\text{M}=-40$~$\mu$C/cm$^2$ to $\sigma_\text{M}=40$~$\mu$C/cm$^2$, selected to explore a physical range of electrode potentials~\cite{suppmat}.

Upon linear-response predictions of $U-U_\text{PZC}$, the differential capacitance of the interface is computed as $C_\text{diff} = \partial \sigma_\text{M}/\partial U$. 
The resulting C--V profiles are reported in Fig.~\ref{fig:ag-water-fukui} for NaF concentrations that go from 0.005~M to 0.1~M. When compared with the experimental results of Ref.~\cite{Valette1989}, our model can, despite its simplicity, successfully reproduce the characteristic ``camel-shape" of $C_\text{diff}$, retrieving the physical range of electrode potentials around the PZC. In particular, while a fully-quantitative description of Na$^+$ adsorption at negative voltages suffers from the lack of an explicit solvation structure~\cite{Karttunen2006, Landstorfer2016,Li2022}, a remarkable accuracy is achieved in positioning the capacitance peak at positive voltages, related to the adsorption of F$^-$. Importantly, this would not be possible by relying on a purely classical model, as the derived potential step variations are known to substantially overshoot the expected values of $U-U_\text{PZC}$. This shortcoming of classical approaches is
commonly attributed to the lack of electronic spillover~\cite{Willard2009,Scalfi2020thomas-fermi, Jeanmairet2022,Nair2025,Wang2025}. Indeed, our results indicate that predicting the Fermi-energy variations via a Fukui-function based response
naturally filters the classical potential variation in the critical interfacial region~\cite{suppmat}, producing electrode potentials that reflect the missing quantum spillover physics.
 
The derived finite-field pathway for the calculation of Fukui functions promotes conceptual-DFT principles, commonly adopted as mostly heuristic rationalization tools, to a quantitative linear-response framework capable of determining Fermi energy variations at metallic interfaces. {At a fundamental level, our results reveal a direct physical connection between electrostatic screening in electronic conductors and the interfacial potential-step variations via a local response function. This aspect is in stark contrast with traditional linear-response formulations of DFT~\cite{keldysh1989dielectric,giuliani2005quantum,martin2016interacting}, where the effect of interfacial screening on the Fermi-level shift can only be indirectly computed and requires going through the nonlocal electronic susceptibility.} 
Perturbative DFT methods are in fact rarely adopted to determine the response to arbitrary external potentials~\cite{Baroni2001}; instead, different versions of the theory are derived to address specific vibrational and electric-field perturbations~\cite{degironcoli1995,shang+njp2018,Zabalo2024}.  In this context, an insightful relationship between local and nonlocal linear responses can be established upon realizing the formal correspondence between the Fukui function and the inverse dielectric function evaluated at any equipotential point in the metallic bulk~\cite{suppmat}, i.e., $f(\mathbf{r}) \simeq \epsilon^{-1}(\mathbf{r},\mathbf{r}_{\mathrm{bulk}})$. 

It should be noted that many computational strategies applied to metallic interfaces fix the Fermi level to a prescribed value, thereby measuring the interfacial step variation of the electrostatic potential exclusively through the thermally-averaged asymptotic value in the electronically insulating region. This is, for instance, the case encountered in extended Lagrangian classical simulations under constant potential~\cite{Siepmann1995, Reed2007,Dufils2019}, as well as in grandcanonical formulations of density functional theory~\cite{horm2019jcp,Melander2019}. In the latter case, the idea behind our method can be similarly adopted to determine how an external perturbation affects the instantaneous surface charge $Q$ at the fixed value of $E_\text{F}$, as appears through the relevant Legendre transform. However, the underlying physical problem is equivalent in complexity to that of fixing the number of electrons, and the choice between canonical and grandcanonical descriptions remains a matter of practical convenience~\cite{Hormann2024}. 

Beyond the calculation of electrode potentials, the acquired ability to provide a  quantitative estimate of work-function changes to local electrostatic perturbations could help rationalizing preadsorption onset potentials in electrocatalytic processes~\cite{Deshlahra2009,Che2017,Che2018}, as well as in understanding the role of surface dipoles in molecular chemisorptions, e.g., self-assembled monolayers, relevant for tuning the properties of organic electronic devices~\cite{Campbell1996,Rusu2006,Rusu2006prb,Hohman2009,Hofmann2010,Hofmann2017,Tsvetanova2020}. 
Furthermore, the theory could have a direct impact in the calculation of tunneling currents under near-contact regimes~\cite{Blanco2004,Krej2017,Litman2023,brezina2025}, where predicting locally induced variations of the Fermi level is required to go beyond the standard Tersoff-Hamann approximation~\cite{Tersoff1985,Hofer2003}. 

{Finally, we note that the assumption of perfect electrostatic screening precludes direct application of the theory to insulating materials or metallic structures in the sub-nanometer regime, where the Fermi-energy response to variations of the external potential is not necessarily localized. Extending the framework to treat weak screening regimes and nonplanar geometries represents an interesting direction for future work.} 

\section{Acknowledgments} The authors are grateful to Pietro Ballone, Federico Grasselli, Mariana Rossi, Benjamin Rotenberg, and Mathieu Salanne for providing fruitful comments on an early version of the manuscript.
The authors acknowledge HPC resources granted by GENCI, France (resources of CINES, Grant No. A0170910463).

\section{Data availability}

The data that support the findings of this article are publicly available~\cite{zenodo}.

\end{document}